\documentclass[12pt]{article}
\usepackage[dvips]{color}
\usepackage{epsfig}
\usepackage{amsmath}
\usepackage{graphicx}
\def\Box{\hbox{$\rlap{$\sqcup$}\sqcap$}}
\begin{document}
\title{\bf  String Inspired Quintom Model with Non-minimally Coupled Modified Gravity }
\author{J. Sadeghi $^{a}$\thanks{Email: pouriya@ipm.ir}\hspace{1mm}
,M. R. Setare $^{b,c}$ \thanks{Email: rezakord@ipm.ir}\hspace{1mm},
 A. Banijamali $^{a}$\thanks{Email: abanijamali@umz.ac.ir}\hspace{1mm}   \\
$^a$ {\small {\em  Sciences Faculty, Department of Physics, Mazandaran University,}}\\
{\small {\em P .O .Box 47415-416, Babolsar, Iran}}\\
$^{b}${\small {\em Department of Science, Payame Noor University,
Bijar, Iran }}\\$^{c}${\small {Research Institute for Astronomy and
Astrophysics of Maragha, P. O. Box 55134-441, Maragha, Iran. }}}

\maketitle

\begin{abstract}
In this paper we consider a quintom model of dark energy with
non-minimal coupling between scalar field and modified gravity which
is known $f(R)$ gravity. The Lagrangian for scalar field has been
inspired by tachyonic Lagrangian in string theory. Then we obtain
the equation of state (EoS), and the condition required for the
model parameters when $\omega$ crosses over $-1$. This model shows
that for having $\omega$ across over -1, one doesn't need to add
some higher derivative operator in the tachyonic part of action (
the way that usually used to obtain crossing of the phantom divide
line for EoS parameter ).
\\


\end{abstract}
\newpage
\section{Introduction}
Nowadays it is strongly believed that the universe is experiencing
an accelerated expansion, and this is supported by many cosmological
observations, such as SNe Ia \cite{1}, WMAP \cite{2}, SDSS \cite{3}
and X-ray \cite{4}. There are two ways to explain the current
accelerated expansion of the universe.  The first one is to
introduce some unknown matter, which is called dark energy in the
framework of general relativity.
 Although the nature and origin of
dark energy could perhaps understood by a fundamental underlying
theory unknown up to now, physicists can still propose some
paradigms to describe it.  The most obvious theoretical candidate
for dark energy is the cosmological constant \cite{5,Weinberg89,6}
which has the equation of state $\omega =-1$. However, it leads to
the two known difficulties \cite{7}, namely the ``fine-tuning''
problem (why is the current vacuum energy density so small), and the
``cosmic coincidence'' one (why are the densities of vacuum energy
and dark matter nearly equal today since they  scale very
differently during the expansion history). In the other side the
analysis of the properties of dark energy from recent observations
mildly favor models with $\omega$ crossing -1 (phantom divide line)
in the near past \cite{c6}. In the framework of general relativity,
the crossing of the phantom divide line has been realized in the
literatures in different approaches, such as scalar tensor theories
with non-minimal coupling between scalar field and curvature
\cite{c7}, two scalar field models \cite{c8} string-inspired models
\cite{c9} and so on. In this framework the general belief is that
the crossing of the phantom divide is not admissible in simple
minimally coupled models and its explanation requires models with
non-minimal coupling between scalar field and gravity \cite{c10}.
 As it was indicated in the
literature \cite{11}, the consideration of the combination of
quintessence \cite{quintessence} and phantom \cite{phantom} in a
unified model, leads to the fulfillment of the aforementioned
transition through the $w=-1$ divide. This model, dubbed quintom,
can produce a better fit to the observational data.
\\
The second way to account for the current accelerated expansion of
the universe is to modify the gravitational theory and in the
simplest case replace $R$ with $f(R)$ in the action which is well
known as $f(R)$ gravity. Here $f(R)$ is an arbitrary function of
scalar curvature (for recent reviews see \cite{{c11},{c12}}).
\\
Although there are some works with related subjects on crossing of
the phantom divide line in the framework of modified gravity
\cite{{c11},{c13},{c14}}, but Ref.\cite{c15} was the first paper
that has investigated a modified gravity model realizing $\omega$
across -1. The authors of Ref.\cite{c15} have shown an explicit
model of modified gravity in which a crossing of the phantom divide
can occur and relation between scalar field theories with property
of $\omega$ crossing -1 and the corresponding modified gravity
theories have been investigated.\\
 In the present paper we would
like to explore the consequence of possibility of a crossing of the
phantom divide line in modified gravity non-minimally coupled with
scalar field. In this model , the tachyon field in the world volume
theory of the open string stretched between a D-brane and an
anti-D-brane or a non-BPS D-brane plays the role of scalar field
\cite{{c16},{c18}}. Although crossing of the phantom divide line can
be realized by using tachyonic matter, but it has been shown that
one needs to add a higher derivative operator in the action and the
extra term plays a important role for having $\omega$ across over
-1. We will show that if we consider non-minimal coupling between
modified gravity and tachyon matter the modification of tachyon
action is not necessary and crossing of the phantom divide line can
occur. An outline of this paper is as follows. In section 2 we
introduce action for tachyon non-minimally coupled to modified
gravity .  In order to discuss the equation of state we derive the
corresponding energy density and pressure for this model. By solving
this equation we obtain the conditions required for the $\omega$
across -1. Section 3 is devoted to discussion of our results.

\section{Non-minimally coupled modified gravity with tachyon field}
We consider the following action for non-minimally coupled $f(R)$
gravity and Born-Infeld type action for tachyon field,
\begin{equation}
S=\int d^{4}x \sqrt{-g} \left[\frac{M_{P}^{2}}{2}f(R)h(\phi) -
V(\phi)\sqrt{1+\alpha'\nabla_{\mu}\phi\nabla^{\mu}\phi}\right],
\end{equation}
where $h(\phi)$ is a function of the tachyon $\phi$ and corresponds
to the non-minimal coupling factor. Here $V(\phi)$ is the tachyon
potential which is bounded and reaching its minimum asymptotically.
$M_{P}=\frac{1}{\sqrt{8\pi G}}$ is reduced
Planck mass.\\
The equation of motion of the scalar field is as follows;
\begin{eqnarray}
\alpha'
\nabla_{\mu}(\frac{V(\phi)\nabla^{\mu}\phi}{u})-V_{\phi}(\phi)u+\frac{M_{P}^{2}}{2}f(R)h_{\phi}(\phi)=0,
\end{eqnarray}
where\\
 $$V_{\phi}(\phi)=\frac{dV(\phi)}{d\phi},\hspace{.5cm}
h_{\phi}(\phi)=\frac{dh(\phi)}{d\phi}, \hspace{.5cm}
u=\sqrt{1+\alpha'\nabla_{\mu}\phi\nabla^{\mu}\phi}.$$\\
By using the definition of the energy momentum tensor,
\begin{eqnarray}
\delta_{g_{\mu\nu}}S=-\int d^{4}x\frac{\sqrt{-g}}{2}T^{\mu\nu}\delta
g_{\mu\nu},
\end{eqnarray}
one can obtain the result as;
$$T_{\mu\nu}=g_{\mu\nu}\left(\frac{M_{P}^{2}}{2}f(R)h(\phi)-V(\phi)u\right)$$
\begin{eqnarray}
-M_{P}^{2}\left[f'(R)h(\phi)R_{\mu\nu}+
(g_{\mu\nu}\Box-\nabla_{\mu}\nabla_{\nu})f'(R)h(\phi)\right]+\frac{\alpha'V(\phi)\nabla_{\mu}\phi\nabla^{\nu}\phi}{u},
\end{eqnarray}
where $f'(R)=\frac{df(R)}{dR}$.\\
 For a flat Friedman- Robertson- Walker
(FRW) spacetime with the metric
\begin{eqnarray}
ds^{2}=-dt^{2}+a^{2}(t)(dr^{2}+r^{2}d\Omega^{2})
\end{eqnarray}
 and a homogenous scalar field $\phi$,
the equation of motion can be written by the following  equation,
\begin{eqnarray}
\ddot{\phi}+3H\dot{\phi}=
\frac{\alpha'\dot{\phi}^{2}\ddot{\phi}}{(1-\alpha'\dot{\phi}^{2})}-\frac{V_{\phi}(\phi)\dot{\phi}^{2}}{V(\phi)}-
\frac{M_{P}^{2}}{2\alpha'}\frac{f(R)
h_{\phi}(\phi)}{V(\phi)}(1-\alpha'\dot{\phi}^{2})^{\frac{1}{2}},
\end{eqnarray}
while energy density, pressure and Friedman equation are
\begin{eqnarray}
\rho=\frac{V(\phi)}{\sqrt{1-\alpha'\dot{\phi}^{2}}}-\frac{M_{P}^{2}}{2}\left[f(R)h(\phi)+6H\frac{\partial}{\partial
t}(f'(R)h(\phi))-6f'(R)h(\phi)(\dot{H}+H^{2})\right],
\end{eqnarray}
\hspace{.7cm}$p=-V(\phi)\sqrt{1-\alpha'\dot{\phi}^{2}}$
\begin{eqnarray}
+\frac{M_{P}^{2}}{2}\left[f(R)h(\phi)+4H\frac{\partial}{\partial
t}(f'(R)h(\phi))+2\frac{\partial^{2}}{\partial
t^{2}}(f'(R)h(\phi))-2f'(R)h(\phi)(\dot{H}+3H^{2})\right],
\end{eqnarray}

\begin{eqnarray}
H^{2}=\frac{1}{3M_{P}^{2}}\frac{V(\phi)}{\sqrt{1-\alpha'\dot{\phi}^{2}}}-\frac{1}{6}\left[f(R)h(\phi)+6H\frac{\partial}{\partial
t}(f'(R)h(\phi))-6f'(R)h(\phi)(\dot{H}+H^{2})\right].
\end{eqnarray}
Where we have used the following components of $R_{\mu\nu}$ in FRW
spacetime,
\begin{eqnarray}
R_{00}=-3(\dot{H}+H^{2}),\hspace{.5cm} R_{0i}=0,\hspace{.5cm}
R_{ij}=(\dot{H}+3H^{2})g_{ij}.
\end{eqnarray}
$H=\frac{\dot{a}}{a}$ is the Hubble parameter and $a$ is the scale
factor.\\
One can express energy density and pressure in terms of derivatives
of $f(R)$ and $h(\phi)$ with respect to their arguments and time
derivatives of $R$ as follows,
\begin{eqnarray}
\rho&=&\frac{V(\phi)}{\sqrt{1-\alpha'\dot{\phi}^{2}}}\notag\\
&-&\frac{M_{P}^{2}}{2}\left[f(R)h(\phi)-6f'(R)h(\phi)(\dot{H}+H^{2})+6H\dot{R}f''(R)
h(\phi)+6H\dot{\phi}f'(R)h_{\phi}(\phi)\right],
\end{eqnarray}

$$p=-V(\phi)\sqrt{1-\alpha'\dot{\phi}^{2}}
+\frac{M_{P}^{2}}{2}\Big[f(R)h(\phi)-2f'(R)h(\phi)(\dot{H}+3H^{2})+4H\dot{R}f''(R)
h(\phi)$$
$$
+4H\dot{\phi}f'(R)h_{\phi}(\phi)+2\dot{R}^{2}f'''(R)h(\phi)+2\ddot{R}f''(R)h(\phi)+4\dot{\phi}\dot{R}f''(R)h_{\phi}(\phi)
$$
\begin{eqnarray}
+2\ddot{\phi}f'(R)h_{\phi}(\phi)
+2\dot{\phi}^{2}f'(R)h_{\phi\phi}(\phi)\Big],
\end{eqnarray}
 We now study the cosmological evolution of equation of state for
the present model. The equation of state is $p=\omega \rho$. To
explore the possibility of the $\omega$ across -1, we have to check
$\frac{d}{dt}(\rho+p)\neq0$ when $\omega\longrightarrow-1$. \\
From equations (11) and (12) one can obtain the following
expression,
$$\rho+p=\frac{\alpha'V(\phi)\dot{\phi}^{2}}{\sqrt{1-\alpha'\dot{\phi}^{2}}}$$
$$+\frac{M_{P}^{2}}{2}\Big[4\dot{H}f'(R)h(\phi)-2H\dot{R}f''(R)h(\phi)+2\dot{R}^{2}f'''(R)h(\phi)+
2\ddot{R}f''(R)h(\phi)$$
\begin{eqnarray}
+4\dot{\phi}\dot{R}f''(R)h_{\phi}(\phi)+2\ddot{\phi}f'(R)h_{\phi}(\phi)
-2H\dot{\phi}f'(R)h_{\phi}(\phi)++2\dot{\phi}^{2}f'(R)h_{\phi\phi}(\phi)\Big].
\end{eqnarray}
Since $\rho+p=(1+\omega)\rho$,  if we assume $\dot{\phi}=0$ when
$\omega\longrightarrow-1$ , the following condition take place,

$$4H\dot{H}f'(R)h(\phi)-2H^{2}\dot{R}f''(R)h(\phi)=$$
\begin{eqnarray}
-2H\Big(\dot{R}^{2}f'''(R)h(\phi)+
\ddot{R}f''(R)h(\phi)+\ddot{\phi}f'(R)h_{\phi}(\phi)\Big).
\end{eqnarray}
By using above condition as well as $\dot{\phi}=0$, when $\omega$
crosses -1, one can obtain,
$$\frac{d}{dt}(\rho+p)\sim$$
$$h(\phi)\Big[\dot{R}f''(R)(\dot{H}-H^{2})+2f'(R)(H\dot{H}+\ddot{H})+\dot{R}^{3}f''''(R)
+3\dot{R}\ddot{R}f'''(R)+\dddot{R}f''(R)\Big]$$
\begin{eqnarray}
+h_{\phi}(\phi)\Big[\dddot{\phi}f'(R)+3\dot{R}\ddot{\phi}f''(R)\Big].
\end{eqnarray}
One can see from (15) that, even if $\dddot{\phi}=0$ and
$\ddot{\phi}=0$, crossing -1 can be happen. This result is in
contrast with the result of Ref.\cite{c19}, where the authors have
added a term $\phi\Box\phi$  in the square root part of action (1)
and concluded that for having crossing over -1 in case of
$\dot{\phi}=0$, one needs $\ddot{\phi}\neq 0$ and $\frac{d}{dt}\Box
\phi\neq 0$ which means $\dddot{\phi}\neq 0$ when $\omega$ crosses
-1 . Also Ref.\cite{c20} considered a dimension-6 operator
$\Box\phi\Box \phi$ in the Lagrangian of phantom field to propose a
model which admits $\omega$ across over -1. In this note we haven't
added a higher derivative operator in the Lagrangian but we
considered non-minimal coupling between matter and modified gravity.
So, it seems that we don't need to add some terms in square root
part of action (1).
\\

\section{Conclusion}
In this paper, we have considered a crossing of the phantom divide
in modified gravity non-minimally coupled with tachyon matter. As a
result we have shown that instead of modification in tachyonic
square root action, the non-minimally coupled $f(R)$ gravity can
play an important role to realize EoS across over -1.\\
We assumed when crossing over -1 occur, $\dot{\phi}=0$ and concluded
that even if $\ddot{\phi}=0$ and $\dddot{\phi}=0$, our model can
admit crossing of the phantom divide line.  In this model we have
shown that the  modification of tachyon Lagrangian with higher
derivative operator in Ref \cite{c19} is  not  require for  crossing
over $-1$ for equation of state. It will be interesting to examine
this model for the special potential $V(\phi)$ ,  $h(\phi)$ and
$f(R)$ in Refs. \cite{{c21},{c22}}.
\section{Acknowledgment}
The work of M. R. Setare has been supported financially by Research
Institute for Astronomy and Astrophysics of Maragha, Iran.

\end{document}